\documentclass[conference,compsoc]{IEEEtran}
\ifCLASSOPTIONcompsoc
  \usepackage[nocompress]{cite}
\else
  \usepackage{cite}
\fi
\usepackage{xspace}
\newcommand{\bl}[1]{\ensuremath{\boldsymbol{#1}}\xspace}
\newcommand{\samp}[3][1]{\ensuremath{\cubra{#3_{#1},\dots,#3_{#2}}}\xspace}
\newcommand{\is}{\emph{in silico}\xspace}

\usepackage{mathtools}
\DeclarePairedDelimiterX\pafun[2](){#1\nonscript\>\delimsize\vert\nonscript\> \mathopen{}#2}
\DeclarePairedDelimiterX\papro[2][]{#1\nonscript\>\delimsize\vert\nonscript\> \mathopen{}#2}
\DeclarePairedDelimiterX\paden[3](){#1\nonscript\>\delimsize\vert\nonscript\> \mathopen{}#2,#3}
\DeclarePairedDelimiterX\pastu[4](){#1\nonscript\>\delimsize\vert\nonscript\> \mathopen{}#2,#3,#4}
\DeclarePairedDelimiterX\palik[2](){#1\nonscript\>;\> \mathopen{}#2}
\DeclarePairedDelimiterX{\cb}[1]\{\}{#1}
\DeclarePairedDelimiterX{\rb}[1](){#1}
\DeclarePairedDelimiterX{\sqb}[1][]{#1}
\DeclarePairedDelimiter{\abs}{\lvert}{\rvert} 

\newcommand{\sqbra}[1]{\sqb*{#1}\xspace}
\newcommand{\paren}[1]{\rb*{#1}\xspace}
\newcommand{\cubra}[1]{\cb*{#1}\xspace}

\DeclareMathOperator{\cor}{Cor}
\DeclareMathOperator{\cov}{Cov} 

\newcommand{\nork}[4][\bl x]{\ensuremath{\mathrm N_{#4} \paden*{#1}{#2}{#3}}\xspace}

\ifCLASSINFOpdf
\else
\fi
\usepackage{array}
\begin{document}
%
\title{Generation of digital patients for the simulation of tuberculosis with UISS-TB}

\author{
\IEEEauthorblockN{Marzio Pennisi\IEEEauthorrefmark{3},Miguel A. Juárez\IEEEauthorrefmark{1},Giulia Russo\IEEEauthorrefmark{2},Marco Viceconti\IEEEauthorrefmark{4},Francesco Pappalardo\IEEEauthorrefmark{2}}
\IEEEauthorblockA{\IEEEauthorrefmark{1} School of Mathematics and Statistics\\
University of Sheffield\\
S3 7RH, Sheffield, UK\\
Email: m.juarez@sheffield.ac.uk}

\IEEEauthorblockA{\IEEEauthorrefmark{2} \textit{Department of Drug Sciences},\\\textit{University of Catania},\\
Catania, Italy\\
\{giulia.russo;francesco.pappalardo\}@unict.it}

\IEEEauthorblockA{\IEEEauthorrefmark{3}\textit{Department of Mathematics and Computer Science},\\\textit{University of Catania},\\
Catania, Italy\\
mpennisi@dmi.unict.it}

\IEEEauthorblockA{\IEEEauthorrefmark{4}\textit{Department of Industrial Engineering},\\\textit{University of Bologna},\\
Bologna, Italy\\
marco.viceconti@unibo.it}

}


%


\maketitle

\begin{abstract}
 EC funded STriTuVaD project aims to test, through a phase IIb clinical trial, two of the most advanced therapeutic vaccines against tuberculosis. In parallel, we have extended the Universal Immune System Simulator to include all relevant determinants of such clinical trial, to establish its predictive accuracy against the individual patients recruited in the trial, to use it to generate digital patients and predict their response to the HRT being tested, and  to combine them to the observations made on physical patients using a new in silico-augmented clinical trial approach that uses a Bayesian adaptive design. This approach, where found effective could drastically reduce the cost of innovation in this critical sector of public healthcare.
One of the most challenging task is to develop a methodology to reproduce biological diversity of the subjects that have to be simulated, i.e., provide an appropriate strategy for the generation of libraries of digital patients. This has been achieved through the the creation of the initial immune system repertoire in a stochastic way, and though the identification of a “vector of features” that combines both biological and pathophysiological parameters that personalize the digital patient to reproduce the physiology and the pathophysiology of the subject.

\end{abstract}


%
\IEEEpeerreviewmaketitle

\section{Introduction}

Tuberculosis (TB) represents one of the world’s deadliest diseases: one third of the world’s population, mostly in developing countries, is infected with TB. But TB is becoming again very dangerous also for developed countries, due to the increased mobility of the world population, and the appearance of several new bacterial strains that are multi-drug resistant (MDR). There is now a growing awareness that TB can be effectively fought only working globally, starting from countries like India, where the
infection is endemic. 

Once a person presents the active disease the most critical issue is the current duration of the therapy, because of the high costs it involves, the increased chances of non-compliance (which increase the probability of developing
an MDR strain), and the time the patient is still infectious to others. One exciting possibility to shorten the duration of the therapy is represented by new host-reaction therapies (HRT) as a coadjuvant of the antibiotic therapy. The endpoints in the clinical trials for HRTs are time to sputum culture conversion, and incidence of recurrence. While for the first it is in some cases possible to have a statistically powered evidence for efficacy in a phase II clinical trial, recurrence almost always requires a phase III clinical trial with thousands of patients involved, and huge costs. 

In the STriTuVaD multidisciplinary consortium we are going to test, through a phase IIb clinical trial, two of the most advanced therapeutic vaccines against drug sensistive tuberculosis (DS-TB) and multi-drug resistant tuberculosis (MDR-TB) i.e., RUTI vaccine, provided by Archivel Farma S.L (Spain) and ID93+GLA-SE vaccine, provided by Infectious Disease Research Institute (US).

In parallel we  extend the Universal Immune System Simulator to include all relevant determinants of such clinical trial, establish its predictive accuracy against the individual patients recruited in the trial, use it to generate digital patients and predict their response to the HRT being tested, and combine them to the observations made on physical patients using a new in silico-augmented clinical trial approach that uses a Bayesian adaptive design. This approach, where found effective could drastically reduce the cost of innovation in this critical sector of public healthcare.

To reproduce biological diversity of the subjects that have to be simulated, an appropriate strategy for the generation of libraries of digital patients has been developed. This has been achieved through the identification of a “vector of features” that combines both biological and pathophysiological parameters that personalize the digital patient.

In this paper, after a brief  recall about UISS and its extension to tuberculosis (sect. \ref{uiss}), we sketch the strategy we adopt to generate the cohort of digital patients(sect. \ref{sec:vp}), and we  show some preliminary results about the dynamics of MTB on a subset of  these patients. (sect. \ref{sec:res}).

\section{Extension of the UISS computational framework to reproduce TB}
\label{uiss}
We will briefly describe here the UISS computational framwork and its extension to model tuberculosis, UISS-TB. The interested reader can find more details about UISS-TB in \cite{Pennisi2019abc}.
\subsection{Introduction to the UISS modeling framwork}
\label{sect:uiss}
UISS is a multi-agent  framework for the simulation of the immune system dynamics that can be extended to reproduce specific diseases and related treatments.
Differently from classical top-down approaches, in which mean behaviors are studied by means of differential equations as presented in \cite{pmid27185314,Castiglione2014a,Pappalardo2014b}, in agent based models and multi-agent systems entities are followed individually, and global nonlinear behaviors arise as the sum of individual behaviors. UISS has been developed as a multi-scale computer simulator of the immune system, as it takes into account both cellular and molecular entities and processes.  

UISS has a long track record of successful stories that include, among others, its use for modelling the effects of a vaccine against the onset of mammary carcinoma \cite{Pappalardo2006g,Palladini2010a} and consequent lung metastases \cite{Pennisi2010b}, for the initial stages of atherosclerosis \cite{pappalardo2008c}, for melanoma \cite{pappalardo2011a}, and more recently, for the study of Multiple Sclerosis \cite{Pennisi2015,Pappalardo2018a} and for testing the  efficacy of  citrus-derived adjuvants for influenza vaccines and human papilloma virus \cite{Pappalardo2016a,Pennisi2017a}.

We then extended UISS to include all the MTB dynamics along with the artificial immunity induced by vaccination strategies as presented in \cite{Pennisi2019abc}.

Finally, to depict individual diversity, a vector of features has been identified. It combines both biological and pathophysiological parameters that personalize the digital patient to reproduce the physiology and the pathophysiology of the subject. In particular, the digital patient model defines a specific patient through 26 features: Drug Sensitive (DS)/Multi-drug resistant (MDR); Bacteria Load (BL) in the sputum; MTB strain; CD4 Th1; CD4 Th2; IgG titers; CD8 T cells; IL-1; IL-2; IL-10; IL-12; IL-17; IL-23; IFN Type I: IFN$\gamma$; TNF$\alpha$; TGF$\beta$; LXA4; PGE2; Chemokines; Vitamin D; HLA-1; HLA-2; FoxP3; Age; BMI. 
The list of parameters, together with the relative range, is presented in table \ref{tab:parameters}. 

\begin{table}
	\centering
	{%
\begin{tabular}{ llc }
\hline
	Feature Description            & Type and Range \\[4pt] \hline \hline
	Drug Sensitive status    & Y/N (MDR otherwise) \\[1pt]
	Bacterial load   in the sputum   & [50-400] CFU (D)  \\[1pt]
	MTB  virulence    & [0-1] (Adimemsional)  (C) \\[1pt] 
	CD4 T  cell type 1   & approx.1370  cells/$\mu$L (D)  \\[1pt] 
	CD4 T  cell type 2   & approx.1370  cells/$\mu$L (D)  \\[1pt] 
	
	Specific  IgG titers  & [2-8] IgG titer (C) \\[1pt] 
	CD8 T  cell     & approx.560  cells/$\mu$L (D)  \\[1pt]

	Interleukin 1 & [0-10000]  pg/mL  (C)  \\[1pt] 

	Interleukin 2 &[50-1000]  pg/mL  (C)  \\[1pt] 

	  Interleukin 10 &[5-16]   pg/mL  (C)  \\ [1pt]
	  Interleukin 12 &[3-300]  pg/mL  (C)   \\ [1pt]
	  Interleukin 17 & [0-1000]  pg/mL  (C)  \\ [1pt]
	  Interleukin 23 & [0-1000]  pg/mL  (C)  \\ [1pt]
	  Type 1 Interferon &[0-11000]  pg/mL  (C)  \\ [1pt]
	  Interferon-$\gamma$ &[ [6-19] pg/mL  (C)  \\ [1pt]
	    TNF-$\alpha$ &[4-40] pg/mL  (C) \\ [1pt]
	  TFG-$\beta$  &[2-8]  pg/mL  (C) \\ [1pt]

	  LXA4& [0-3] ng/mL (C) \\ [1pt]
	  PGE2&[0-2.2]ng/mL (C) \\ [1pt]

	  General chemokine &[0-20] ng/mL (C) \\ [1pt]

	  Vitamin D &[0-100] ng/mL (C) \\ [1pt]

	HLA-class 1   & [$0-2^{\mbox{NBITSTR}-1}$]  (Adimemsional) (D) \\ [1pt]
	HLA-class 2  & [$0-2^{\mbox{NBITSTR}-1}$]  (Adimemsional) (D) \\ [1pt]

	FoxP3  cells & approx.60 cells/$\mu$L (D) \\ [1pt]

	Age     & [0-90]  years (D) \\ [1pt]

	Body Mass  Index  & [16-41] $Kg/m^2$   (C)    \\ [1pt]

\end{tabular}
	}
	\vspace{.1cm}
	\caption{Vector of features and relative range used to identify and define digital patients. (D) stands for discrete variable; (C) stands for continuous variable. }
	\label{tab:parameters}
	\vspace{-.2cm}
\end{table}

\section{Generation of Digital Patients: a Bayesian approach}
\label{sec:vp}
\subsection{The UISS-TB input vector}

 The UISS-TB model defines a specific patient through a vector of 26 features as described in section \ref{sect:uiss-tb}:
	
In order to create an \is patient, one needs to provide a single value for each one of 1--26. These values could be taken from individual physical patients; however, if a cohort of digital patients is to be produced, one should have a mechanism for producing as many different input vectors as needed, that are biological/physiological plausible. Formally, this requires the characterisation of the joint distribution of the inputs in the population.  We have compiled typical values and standard deviations for each feature, providing a way to generate plausible values for each component at a time.  Proceeding in this way would neglect the biological correlations between features and thus would not guarantee a physiologically plausible input vector.  Hence, we must take into account these correlations.  Given that we have 25 numerical input variables (DS/MDR is a factor), we should specify $25 \times 24/2 = 300$ correlations. Using relevant literature and expert opinion, we have qualified these correlations, determining that all correlations are positive, but the correlation of IL-10 with the rest of the features.

\subsection{Formalising \is profile generation} \label{sub:formalising_is_profile_generation}
In theory, one could elicit the joint distribution of the 25 features, i.e. describe mathematically how each feature relate to each other in a space of 25 dimensions; but this would be not only extremely difficult, but also time consuming and data demanding.  Our approach is to rely on current mathematical biology consensus and use a Gaussian to represent the population distribution. The additional advantage of using this approach will be discussed in the next section.

Formally, we say that the vector \bl x = \samp dx follows a $d$-variate Gaussian distribution with joint probability density function (pdf)
\[
	\nork{\bl\mu}{\Sigma}d = \begin{multlined}[t] (2 \pi)^{-d/2} \abs{\Sigma}^{-1/2} \> \times \\ \exp\sqbra{- \frac 12 \paren{\bl x - \bl \mu}' \Sigma^{-1} \paren{\bl x - \bl\mu} }, \end{multlined}
\]
with mean $\bl \mu = \samp d\mu$ and covariance matrix,
\[
	\Sigma = \paren{\begin{matrix}\sigma^2_{1} & \sigma_{12} & \hdots & \sigma_{1d} \\ \sigma_{21} & \sigma^2_{2} & \hdots & \sigma_{2d} \\ \vdots & \vdots & \ddots & \vdots \\ \sigma_{d1} & \sigma_{d2} & \hdots & \sigma^2_{d} \end{matrix}}
\]
where,
\begin{gather*}
	\cov\paren{x_i, x_j} = \sigma_{ij}
	\shortintertext{related to the correlations by}
	\cor\paren{x_i, x_j} = \rho_{ij} = \frac{\sigma_{ij}}{\sqrt{\sigma^2_i\sigma^2_j}}
\end{gather*}
So, if we are able to elicit a measure of correlation between two inputs, we can calculate their covariance.

The elements in the diagonal, $\sigma^2_i$ are the marginal variances of each element, $x_i$, and $\mu_i$ the corresponding marginal mean.  As mentioned above, we already have compiled a list with these values, so we have elicited values for \bl\mu and the diagonal elements of $\Sigma$, $\sigma^2_i$.

\subsection{Cohort generation} \label{sub:cohort_generation}

Once $\bl \mu$ and $\Sigma$ have been elicited, generating an \is profile is a relatively trivial task: one must sample a point in the25-dimensional space, consistent with \nork{\bl\mu}{\Sigma}d.  But we can exploit the properties of the Gaussian distribution to produce a cohort consistent with some specific characteristics.  Say, for instance, that our target population has a particular range of BL, we would like then to produce digital patients consistent with that specific profile.  Formally, let $x_1$ represent BL and $\bl x_{-1} = \samp[2]{18}x$, the rest of the features; we would like to sample from
\[
	\nork[\bl x_{-1}]{x_1, \bl\mu}{\Sigma}d
\]
i.e. the conditional distribution of the rest of the features, given that BL has a specific value.  This is a standard procedure, which can be readily implemented.  We can go even further and sort the list of features according to either their importance in determining the profile of a patient, or to the precision of their elicited mean, variance and covariance, and then proceed to sample from the conditional distributions, one at a time.

\section{Preliminary results}
\label{sec:res}
To test the approach presented in sect. \ref{sec:vp} we created an R script for the generation of  digital patents. In table \ref{tab:vp} we report  30 generated digital patients using the aforementioned approach. All the patients have been then simulated using UISS-TB.

\begin{table*}
	\centering
		\setlength\tabcolsep{1.5pt} 
\begin{scriptsize}
	\begin{tabular}{  l | l | l | l | l | l | l | l | l | l | l | l | l | l | l | l | l | l | l | l | l | l | l | l | l  }

	N.Ag & strain & Th1 & Th2 & IgG & TC & IL1 & IL2 & IL10 & IL12 & IL17 & IL23 & IFN1 & IFNG & TNF & TGFB & LXA4 & PGE2 & Chem & VD & MHC1 & MHC2 & Treg & Age & BMI \\ \hline\hline
	189 & 0.7064& 43 & 46 & 4.53 & 50 & 5061.45 & 555.26 & 14.4 & 166.02 & 534.23 & 516.91 & 5534.17 & 21.86 & 13.02 & 7.53 & 2.96 & 0.89 & 9.15 & 47.43 & 2029 & 2077 & 57 & 39 & 30.72 \\ 
	249 & 0.4037 & 49 & 52 & 4.91 & 52 & 5074.84 & 532.75 & 14.95 & 186.13 & 515.34 & 510.78 & 5443.04 & 18.69 & 17.89 & 3.82 & 2.61 & 0.82 & 11.88 & 55.15 & 2047 & 2061 & 56 & 46 & 32.57 \\ 
	187 & 0.5344 & 63 & 53 & 7.09 & 51 & 4993.29 & 514.32& 15.59 & 158.93 & 510.4 & 522.28 & 5499.65 & 20.16 & 19.5 & 6.83 & 2.92 & 748 & 11.78 & 51.74 & 2067 & 2036 & 51 & 47 & 32.34\\ 
	253 & 0.1497 & 54 & 62 & 4.72 & 44 & 4986.72 & 515.62 & 18.74 & 138.36& 524.74 & 512.97 & 5445.44 & 20.26 & 10.84 & 8.1 & 1.98 & 0.3 & 7.01 & 50.09 & 2021 & 2014 & 47 & 47 & 28.52 \\ 
	175 & 0.2359 & 51 & 43 & 4.74 & 51 & 4925.43 & 509.98 & 18.89 & 140.16 & 500.57 & 475.85 & 5514.87 & 20.29 & 11.76 & 5.66 & 2.45 & 0.41 & 12.16 & 43.49 & 2077 & 2047 & 50 & 47 & 27.82 \\ 
	229 & 0.8574 & 56 & 43 & 6.64 & 48 & 4954.02 & 538.34 & 17.48 & 151.19 & 495.06 & 484.93 & 5484.89 & 21.54 & 13.91 & 3.02 & 2.27& 0.34 & 8.61 & 49.42 & 2072 & 2044 & 46 & 47 & 27.29 \\ 
	221 & 0.7262& 51 & 54 & 4.92 & 46 & 4975.01 & 507.1 & 16.17 & 154.19 & 512.23 & 485.33 & 5477.36 & 18.61 & 15.8 & 5.49 & 2.41 & 0.35 & 8.4 & 48.88 & 2034 & 1990 & 42 & 49 & 29.29 \\ 
	199 & 0.2529 & 47 & 56 & 3.67 & 54 & 5019.55 & 515.44 & 17.39 & 150.84 & 502.38 & 460.19 & 5458.05 & 19.5 & 9.6 & 5.44 & 2.37 & 1.07 & 10.53 & 45 & 2069 & 2029 & 46 & 36 & 30.09 \\ 
	158 & 0.341& 44 & 53 & 4.05 & 59 & 5007.74 & 542.11 & 20.79 & 145.03 & 498.41 & 479.92 & 5460.33 & 20.79 & 12.38 & 6.77 & 2.68 & 0.71 & 9.75 & 51.58 & 2066 & 1992 & 52 & 42 & 29.61 \\ 
	191 & 0.2069 & 40 & 45 & 3.64 & 43 & 4981.71 & 519.91 & 18.01 & 138.63 & 494.85 & 503.23 & 5395.49 & 21.54 & 7.92 & 5.73 & 2 & 0.64 & 8.36 & 41.33 & 2024 & 2076 & 37 & 41 & 26.83 \\ 
	328 & 0.56& 48 & 40 & 4.32 & 50 & 5018.13 & 513.28 & 17.84 & 144.18 & 501.32 & 505.75 & 5519.12 & 20.22 & 12.06 & 5.87 & 2.24 & 0.53 & 7.9 & 58.25 & 2033 & 2065 & 40 & 46 & 25.81 \\ 
	252 & 0.5409 & 35 & 53 & 5.44 & 43 & 4973.13 & 496.53 & 18.62 & 143.75 & 478.83 & 500.98 & 5493.01 & 21.74 & 12.08 & 6.12 & 2.09 & 0.9 & 8.26 & 52.65 & 1991 & 2041 & 49 & 38 & 29.29 \\ 
	261 & 0.496 & 54 & 56 & 6.2 & 56 & 4946.85& 525.38 & 17.88 & 153.03 & 502.42 & 502.62 & 5587.85 & 22.67 & 13.08 & 4.13 & 3.22 & 1.03 & 9.73 & 59.39 & 2077 & 2109 & 51 & 55 & 29.8 \\ 
	273 & 0.8468 & 52 & 49 & 7.31 & 54 & 5035.47 & 550.42 & 16.67 & 173.35 & 505.56 & 504.44 & 5510.07 & 21.51 & 13.46 & 6.78 & 3.05 & 1.4 & 12.71 & 51.72 & 2103 & 2107 & 59 & 54 & 32.99 \\ 
	300 & 0.7984 & 42 & 47 & 6.02 & 50 & 4956.64 & 500.79 & 16.64 & 146.57 & 508.83 & 507.02 & 5487.82 & 21.76 & 12.29 & 1.78 & 3.09 & 1.08 & 11.7 & 46.76 & 2036 & 2037 & 55 & 44 & 25.03 \\ 
	245 & 0.3408& 49 & 43 & 5.74 & 50 & 4916.78 & 522.15 & 21.1 & 134.55 & 481.55 & 497.47 & 5430.43 & 20.29 & 15.08 & 6.08 & 2.94 & 1.03 & 8.74 & 57.86 & 2031 & 2057 & 38 & 42 & 31.76 \\ 
	221 & 0.6764 & 53 & 50 & 6.75 & 45 & 4958.18 & 484.49 & 18.45 & 153.83 & 487.77 & 476.68 & 5491.86 & 21.47 & 13.64 & 3.73 & 2.33 & 1.02 & 8.93 & 53.04 & 2059 & 2032 & 51 & 36 & 29.9 \\ 
	122 & 0.8234 & 54 & 45 & 6.14 & 51 & 4911.57 & 520.65 & 17.55 & 163.81 & 499.58 & 504.62 & 5599.99 & 22.12 & 14.77 & 2.50& 2.24 & 987 & 9.99 & 44.99 & 2039 & 2064 & 53 & 38 & 27.25 \\ 
	242 & 0.816 & 42 & 51 & 4.21 & 51 & 4983.22 & 527.73 & 17.32 & 147.31 & 494.85 & 491.66 & 5440.64 & 19.32 & 12.34 & 5.17 & 2.62 & 0.78 & 9.300& 53.42 & 2041 & 2088 & 48 & 47 & 30.6 \\ 
	313 & 0.5595 & 43 & 51 & 6.19 & 55 & 5001.22 & 514.22 & 17.19 & 155.13 & 499.22 & 488.78 & 5537.17 & 20.38 & 12.86 & 3.14 & 2.27 & 0.33 & 5.61 & 57.84 & 2071 & 1992 & 47 & 42 & 27.05 \\ 
	179 & 2.63E-2 & 42 & 44 & 6.79 & 43 & 5070.49 & 520.66 & 21.95 & 170.17 & 525.5 & 525.2 & 5458.13 & 19.68 & 13.22 & 2.63 & 2.17 & 1.110 & 11.05 & 45.17 & 2020 & 2020 & 50 & 36 & 32.617 \\ 
	345 & 0.3713 & 52 & 44 & 6.2 & 48 & 4975.01 & 541.12 & 17.48 & 155.88 & 507.21 & 502.65 & 5467.4 & 21.17 & 13.29 & 6.82 & 2.76 & 0.87 & 14.8 & 49.63 & 2024 & 2112 & 55 & 56 & 34.29 \\ 
	245 & 0.6411 & 51 & 52 & 5.82 & 57 & 4992.59 & 551.01 & 15.56 & 173.09 & 518.22 & 514.61 & 5569.51 & 22.26 & 13.26 & 8.31 & 2.65 & 1.15 & 13 & 51.33 & 2064 & 2059 & 54 & 49 & 29.1 \\ 
	251 & 0.5810 & 51 & 51 & 2.99 & 49 & 4988.39 & 521.97 & 20.7 & 127.72 & 489.56 & 484.17 & 5438.56 & 18.89 & 12.88 & 3.99 & 2.46 & 0.19 & 5.27 & 53.35 & 2054 & 2069 & 42 & 38 & 24.82 \\ 
	234 & 0.2243 & 46 & 42 & 4.40 & 57 & 5003.68 & 551.15 & 20.72 & 154.75 & 501.44 & 519.77 & 5515.5 & 20.75 & 14.23 & 4.11 & 2.2 & 0.49 & 10.25 & 44.46 & 2021 & 2017 & 45 & 44 & 26.95 \\ 
	186 & 0.4632 & 50 & 47 & 4.36 & 47 & 4869.57 & 551.46 & 20.02 & 138.66 & 496.34 & 500.4 & 5476.01 & 20.67 & 12.43 & 4.49 & 3.1 & 0.74 & 6.99 & 47.49 & 2080 & 2045 & 56 & 45 & 30.6 \\ 
	154 & 0.3493 & 54 & 49 & 6.01 & 44 & 5014.58 & 515.5 & 18.26 & 152.16 & 502.53 & 482.3 & 5576.32 & 20.55 & 13.9 & 5.67 & 2.68 & 0.99 & 10.35 & 56.32 & 2011 & 2028 & 54 & 41 & 27.18 \\ 
	239 & 0.7161 & 58 & 56 & 6.29 & 48 & 4961.32 & 548.38 & 17.89 & 141.16 & 520.83 & 487.15 & 5519.7 & 20.38 & 12.25 & 3.23 & 2.44 & 1.17 & 12.23 & 45.65 & 2057 & 2048 & 54 & 45 & 28.06 \\ 
	208 & 0.498 & 46 & 54 & 4.65 & 52 & 5064.60& 539.2 & 19.04 & 160.08 & 494.1 & 520.26 & 5638.14 & 21.84 & 14.05 & 4.08 & 2.86 & 0.96 & 6.37 & 56.91 & 2062 & 2026 & 45 & 51 & 26.84 \\ 
	209 & 0.5812 & 47 & 57 & 3.77 & 53 & 5118.04 & 530.24 & 17.28 & 149.22 & 501.75 & 492.45 & 5513.76 & 21.82 & 11.72 & 7.06 & 2.69 & 0.4 & 8.76 & 47.44 & 2043 & 2043 & 51 & 44 & 32.81 \\

\end{tabular}
\end{scriptsize}
	
	\vspace{.1cm}
	\caption{Vector of features for 30  Digital Patients generated according the approach described in section \ref{sec:vp}. The column for DS/MDR status is not reported in table.}
	\label{tab:vp}
	\vspace{-.2cm}
\end{table*}

\begin{figure*}
	\centering
	\includegraphics[width=.95\linewidth]{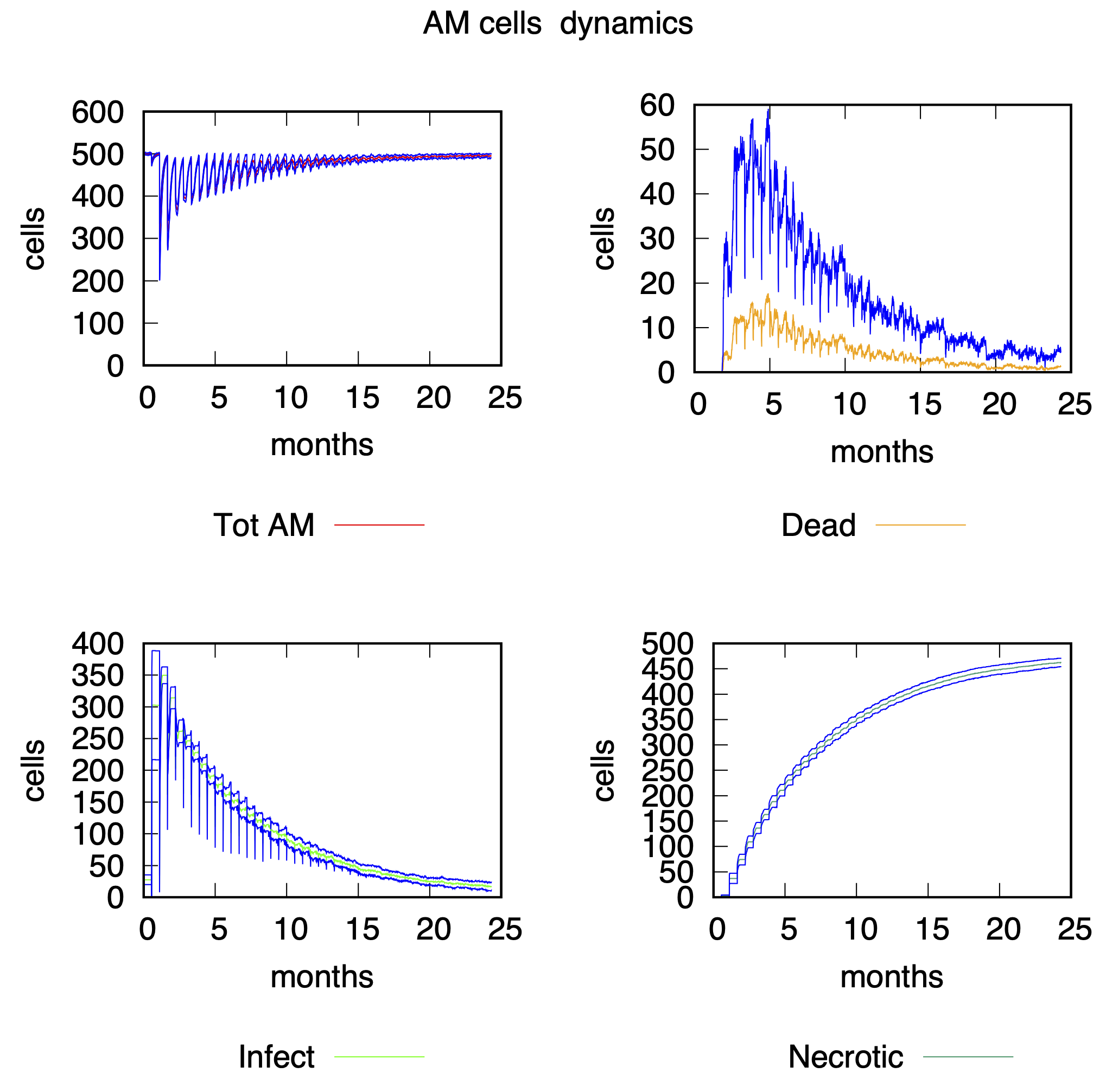}
	\caption{Total (red lines), Dead (yellow lines) Infected (green lines) and Necrotic (dark green lines) Alveolar Macrophages mean behavior for a simulation time of 2 years computed over the 30 random digital patients in table \ref{tab:vp} absence of treatment. Blue lines represent mean +/- SD}.
	\label{fig:notreat}
	\vspace{-.4cm}
\end{figure*}

 In the following figures we show the typical UISS-TB simulation framework when applied to the sample set of digital in silico patients depicted in table \ref{tab:vp}. We show, for each biological entity, both the mean behavior (according to the entities the color line may vary) and the +/- SD (blue lines). We run a total of 30 simulations for untreated digital in silico patients.
In figure  \ref{fig:notreat} it is depicted the dynamics of alveolar macrophages during two phases. The first one is during the active TB phase where both necrotic and MTB-infected populations increase. After the active phase, a latent phase is established, and necrotic alveolar macrophages contribute to typical granuloma formation.

\section{Conclusions and future work}
The set up  an ``in silico'' trial requires that the involved computational model   is able to coherently reproduce the disease dynamics on different individuals. As a consequence of that, it is important to establish a rigorous strategy for the definition of a credible cohort of digital patients. To this end, we presented an approach for creating a set of digital patients whose features can be in line with  those of the real population.

Preliminary results about the execution of UISS-TB on the cohort of digital patients show that the simulator is able to capture the dynamics of  this pathology.

The  next step will be focused on   the generation of reference digital populations to be used as part of the technical validation of UISS-TB. Once the data from the clinical trials will be available, we will regenerate the digital cohorts, and we will use a Bayesian statistical model approach to explore specific use cases, such as that of in silico-augmented clinical trials, where digital and physical patients are combined.

\ifCLASSOPTIONcompsoc
  \section*{Acknowledgments}
\else
  \section*{Acknowledgment}
\fi

Authors of this paper acknowledge support from the STriTuVaD project. The STriTuVaD project has been funded by the European Commission, under the contract H2020-SC1-2017- CNECT-2, No. 777123. The information and views set out in this article are those of the authors and do not necessarily reflect the official opinion of the European Commission. Neither the European Commission institutions and bodies nor any person acting on their behalf may be held responsible for the use which may be made of the information contained therein.



\bibliographystyle{IEEEtran}
\bibliography{Digital_Patients_CMISF_2019_}

\end{document}